\documentclass[12pt,a4paper]{article}

\usepackage[T1]{fontenc}
\usepackage[utf8]{inputenc}
\usepackage[margin=2.5cm]{geometry}
\usepackage{amsmath,amssymb,bm}
\usepackage{graphicx}
\usepackage{placeins}
\usepackage{blkarray}
\usepackage{subcaption}
\usepackage{booktabs}
\usepackage{siunitx}
\usepackage[numbers,sort&compress]{natbib}
\usepackage[colorlinks=true,allcolors=blue]{hyperref}
\usepackage{xcolor}
\usepackage{tikz}
\usetikzlibrary{arrows.meta,positioning,fit,calc}

\graphicspath{{figures/}}

\newcommand{\jphi}{J_\phi}
\newcommand{\jvec}{\bm{j}}
\newcommand{\psivec}{\bm{\psi}}
\newcommand{\yvec}{\bm{y}}
\newcommand{\Ivec}{\bm{I}}
\newcommand{\xvec}{\bm{x}}
\newcommand{\rvec}{\bm{r}}
\newcommand{\bvec}{\bm{b}}
\newcommand{\onevec}{\bm{1}}
\newcommand{\Hmat}{\bm{\mathsf{H}}}
\newcommand{\Gmat}{\bm{\mathsf{G}}}
\newcommand{\Smat}{\bm{\mathsf{S}}}
\newcommand{\Sigmat}{\bm{\Sigma}}

\newcommand{\Mmat}{\bm{\mathsf{M}}}
\newcommand{\Bmat}{\bm{\mathsf{B}}}
\newcommand{\Wmat}{\bm{\mathsf{W}}}
\newcommand{\Amat}{\bm{\mathsf{A}}}
\newcommand{\Imat}{\bm{\mathsf{I}}}

\title{Diffusion prior for KSTAR equilibrium reconstruction under sensor dropout}

\author{Hyungkeun Nam and Jaemin Seo\thanks{Corresponding author: \href{mailto:jseo@cau.ac.kr}{jseo@cau.ac.kr}.}\\[0.5em]
\small Department of Physics, Chung-Ang University\\
\small Seoul, Republic of Korea}

\date{}

\begin{document}
\maketitle

\begin{abstract}
We study equilibrium reconstruction for the Korea Superconducting Tokamak Advanced Research (KSTAR) device under magnetic sensor dropout, with a diffusion model as the prior. Magnetic measurements leave most components of the toroidal current density $\jphi$ undetermined, and sensor loss leaves more of them to the prior. The diffusion model learns only $\jphi$, conditioned on the coil currents, the plasma current and the product of major radius and toroidal field, which do not depend on the dropout. The physics enters through a linear forward operator built from sensor response matrices of the LIUQE code. Because the observations are linear, a variant of decoupled annealing posterior sampling fits them with a closed-form linear correction.
On measured signals of 37 KSTAR shots, we switch off a random fraction (0.0 to 0.9, ten settings) of the 124 magnetic channels in use and compare with LIUQE under the same masks, taking the full-sensor LIUQE reconstruction as the label. Both methods fit the remaining channels to a similar level. The median relative distance of $\jphi$ from the label stays at 3.22--4.07\% for the diffusion reconstruction in all settings, whereas that of LIUQE reaches 11.89\% at dropout 0.9. At dropout 0.5--0.9, the diffusion reconstruction is closer to the label on 27--37 of the 37 shots. At low dropout, LIUQE is closer on most shots. For the derived flux, the diffusion reconstruction is closer on 21--36 shots at dropout 0.5--0.9, and its error at high dropout comes from the vessel current estimate, not from the prior.

\end{abstract}


\section{Introduction}
\label{sec:intro}

Equilibrium reconstruction infers the plasma current distribution of a tokamak from its magnetic diagnostics. These diagnostics measure linear functions of the toroidal current density $\jphi$. If $\jvec$ denotes the vector of $\jphi$ on the reconstruction grid, the poloidal flux $\psivec$ on the grid and its measurements $\yvec$ at the sensors are
\begin{equation}
  \psivec = \Gmat\jvec + \psivec_\mathrm{ext}, \qquad \yvec = \Hmat\jvec + \yvec_\mathrm{ext}.
\end{equation}
Here, the external terms $\psivec_\mathrm{ext}$ and $\yvec_\mathrm{ext}$ are the contributions of the poloidal field (PF) coil currents and of the currents induced in the vacuum vessel, which we call the vessel currents. Bold italic symbols denote vectors, and bold sans-serif symbols denote matrices.

However, external magnetic measurements alone do not determine $\jphi$ uniquely. Substantially different current densities and safety factors $q$ reproduce the same magnetic signals equally well within the measurement errors, as shown for MAST, JET and ITER-like plasmas~\cite{Zaitsev2011}. Choosing one of them therefore requires an assumption from outside the observations, that is, a prior. Standard reconstruction codes such as EFIT~\cite{Lao1985} and LIUQE~\cite{Moret2015} take force balance together with low-order polynomial profiles as this prior. The polynomial profiles reduce the degrees of freedom to a few coefficients, and the components that the observations cannot see are then set by the choice of basis. When sensors are lost, the observations see even fewer directions, and the prior sets a larger share of the answer. Such losses occur in practice through integrator drift~\cite{Strait2008,Joung2020} and coil damage~\cite{Artaserse2019}. Moreover, in next-step devices, radiation induces errors in the magnetic coils~\cite{Vayakis2004}, and the designs of DEMO and SPARC include diagnostic redundancy against sensor failure~\cite{Donne2012,Stewart2023}.

Existing remedies act either on the reconstruction or on the signals. On the reconstruction side, neural networks trained on EFIT reconstructions map the magnetic signals to the equilibrium on DIII-D~\cite{Lao2022}. Such networks are sensitive to missing inputs: on NSTX-U, a network roughly doubled its errors when a single magnetic probe was zeroed~\cite{Wai2022}. On the signal side, one method detects a faulty probe by comparing it with its neighbors~\cite{Nouailletas2012}. On the Korea Superconducting Tokamak Advanced Research (KSTAR) device~\cite{Nam2026}, a few missing probe signals were imputed for real-time reconstruction~\cite{Joung2018}, and a neural network reconstruction later used this imputation with up to nine missing signals~\cite{Joung2023bnn}. These remedies, however, handle only a few missing signals. When many signals are lost, the reconstruction instead needs a prior that fills the larger unobserved share closer to actual equilibria than low-order polynomials do.

\begin{figure}[t]
  \centering
  \includegraphics[width=\textwidth]{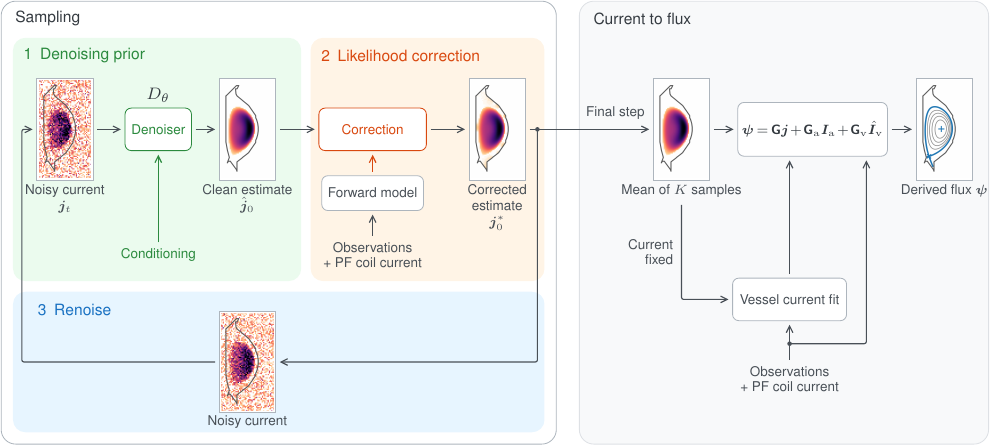}
  \caption{Reconstruction scheme: sampling of $\jphi$ and derivation of the flux $\psivec$. The images are from one reconstruction of shot \#41744 at $t = \qty{8.656}{\second}$ without dropout, and the sampling images show one sample at $\sigma_t = 0.940$.}
  \label{fig:method}
\end{figure}

Diffusion models are a natural candidate for such a prior. They learn the score of the data distribution, $\nabla_{\xvec} \log p(\xvec)$, and generate samples with it~\cite{Song2021}. A trained model therefore draws plausible fields from its training distribution, but these samples ignore any measurement. To serve as a prior in an inverse problem, the sampling must be steered toward the posterior $p(\xvec \mid \yvec) \propto p(\yvec \mid \xvec)\, p(\xvec)$, so that the samples also fit the observations. Existing methods differ in how each step is pulled toward the observations. Guidance adds the gradient of the likelihood to the score at every step and pushes the sample toward the observations~\cite{Chung2024dps}. Projection replaces, at every step, the observed component of the noisy sample with a noised copy of the measurements~\cite{Song2022}. Decoupled annealing posterior sampling (DAPS) instead denoises first and then corrects the clean estimate toward the measurements with a regularized least-squares step, rather than replacing components of the noisy sample. This lets it correct errors from earlier steps, which methods that modify the sample only slightly at each step struggle to do~\cite{Zhang2024}. Our reconstruction uses the same structure, alternating denoising and correction (figure~\ref{fig:method}). In each of these approaches, the prior fills the directions that the observations do not determine.

Where the physics enters is a further design choice. In inverse problems governed by partial differential equations (PDEs), physics enters a diffusion framework in one of two ways. Joint designs learn the PDE coefficients and the solution together and put the physics into the model, through the training loss~\cite{Bastek2025} or through guidance by the PDE residual~\cite{Huang2024}. Decoupled designs, by contrast, learn a prior over the coefficients only and keep the physics in a forward operator~\cite{Wang2023,Shan2025}. Lin et al~\cite{Lin2026} proved that guidance attenuates in joint models when training data are scarce, and therefore chose the decoupled design.

Diffusion priors have already been applied to inverse problems across the physical sciences, including strong gravitational lensing, black hole imaging, fluid data assimilation and inertial confinement fusion~\cite{Adam2022,Feng2023,Rozet2023,Jones2026,Han2026}. On a benchmark of scientific inverse problems, they generally outperformed the established methods of each field when suitable training data were available, but solutions outside the prior distribution were hard to recover~\cite{Zheng2025}. Moreover, their advantage over the baselines held or grew when observations were removed~\cite{Zheng2025,Shan2025}. However, to our knowledge, diffusion priors have not been applied to tokamak equilibrium reconstruction from magnetic diagnostics.

In this work, we transfer the decoupled design to tokamak equilibrium reconstruction from magnetic diagnostics. Our diffusion prior learns only $\jphi$, and the physics stays in the forward operator. Unlike PDE inverse problems, where the forward model must be learned or solved numerically, the magnetic response is linear in $\jphi$ and known exactly. We use the sensor response matrices of LIUQE as the forward operator, so the correction step of DAPS has a closed-form solution. The poloidal flux is not reconstructed but derived from $\jphi$. To evaluate the method, we test it on measured KSTAR signals under random dropout, which switches off a randomly chosen fraction of the 124 channels in use. For each of the ten dropout settings, we compare it with LIUQE run with the same masks. We call the full-sensor LIUQE reconstruction the label, and the distance of $\jphi$ from the label the label distance. In our results, the label distance of the diffusion reconstruction stays flat from dropout 0.0 to 0.9. Moreover, at dropout 0.5--0.9, it is lower than that of LIUQE on 27--37 of the 37 evaluation shots. Likewise, for the derived poloidal flux, the diffusion reconstruction is lower than LIUQE at dropout 0.5--0.9, on 21--36 of the 37 shots. Meanwhile, the fit to the sensors, measured by $\chi^2$, stays at the same level for the two methods. The rest of the paper is organized as follows. Section~\ref{sec:formulation} formulates the problem, section~\ref{sec:method} describes the method and section~\ref{sec:results} gives the results. Section~\ref{sec:discussion} discusses them, and section~\ref{sec:conclusion} concludes.

\section{Problem formulation}
\label{sec:formulation}

\subsection{Observation model}
\label{sec:obs-model}

On the $87 \times 47$ reconstruction grid, the magnetic diagnostics and the poloidal flux are linear in $\jvec$~\cite{Hutchinson2002,Moret2015}. The observations are
\begin{equation}
  \yvec = \Hmat \jvec + \Hmat_\mathrm{a} \Ivec_\mathrm{a} + \Hmat_\mathrm{v} \Ivec_\mathrm{v},
  \label{eq:obs}
\end{equation}
where $\Ivec_\mathrm{a}$ are the measured poloidal field (PF) coil currents and $\Ivec_\mathrm{v}$ are the unmeasured vacuum vessel currents. Each column of $\Hmat$ is the response of the sensors to a unit current in one grid cell, and $\Hmat_\mathrm{a}$ and $\Hmat_\mathrm{v}$ are the same for the coil and vessel currents. The subscripts a and v follow the LIUQE notation~\cite{Moret2015}. The poloidal flux per radian on the grid, in \si{\weber\per\radian}, is
\begin{equation}
  \psivec = \Gmat \jvec + \Gmat_\mathrm{a} \Ivec_\mathrm{a} + \Gmat_\mathrm{v} \Ivec_\mathrm{v}.
  \label{eq:psi}
\end{equation}
All six matrices follow from the mutual inductance between coaxial circular filaments (appendix~\ref{app:operator}). Once $\jvec$ and $\Ivec_\mathrm{v}$ are fixed, equation~\eqref{eq:psi} gives $\psivec$.

\begin{figure}[t]
  \centering
  \includegraphics{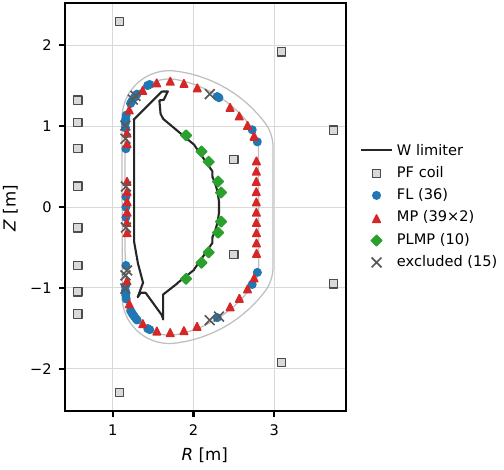}
  \caption{Poloidal cross-section of KSTAR with the flux loops (FL), magnetic probes (MP), poloidal limiter magnetic probes (PLMP), poloidal field (PF) coils, tungsten limiter and vessel walls.}
  \label{fig:geometry}
\end{figure}

The observation rows come from flux loops (FL), magnetic probes (MP), and poloidal limiter magnetic probes (PLMP). KSTAR has 139 observation rows. They consist of 45 FL rows, 84 MP rows from the two components of 42 probes, and 10 PLMP rows. We remove the 15 rows that are permanently missing and use the remaining 124. Figure~\ref{fig:geometry} shows their layout. The vessel currents $\Ivec_\mathrm{v}$ are represented by 104 filaments. Because they are not measured, the reconstruction must eliminate or estimate them.

The likelihood is Gaussian with a standard deviation $\sigma_i$ for each channel $i$. We calibrated $\sigma_i$ from the residuals of the full-sensor LIUQE reconstruction, which we call the label, as $\sigma_i^2 = \langle (y_i - \hat{y}_i^\mathrm{label})^2 \rangle$. Here $\hat{y}_i^\mathrm{label}$ is read off the label flux $\psi^\mathrm{label}$ on the grid. For a flux loop it is $2\pi\psi^\mathrm{label}$ interpolated bilinearly at the loop. For a probe it is the field component along the probe direction, $B_R\cos\theta_i + B_Z\sin\theta_i$, with $B_R = -R^{-1}\partial_Z\psi^\mathrm{label}$ and $B_Z = R^{-1}\partial_R\psi^\mathrm{label}$ taken by central differences at the probe. Because $\psi^\mathrm{label}$ already contains the coil and vessel contributions, this prediction needs no vessel currents. The mean $\langle\cdot\rangle$ runs over the label frames. The calibration frames, \num{101427} frames of 275 shots, exclude the evaluation shots and the shots next to them. Dividing each observation row by its $\sigma_i$ whitens the observations, so that the noise on every row has unit standard deviation. Appendix~\ref{app:operator} gives the whitened coordinates and the elimination of the vessel currents. We measure the fit to the sensors by
\begin{equation}
  \chi^2 = \frac{1}{n} \sum_{i=1}^{n} \left( \frac{\hat{y}_i - y_i}{\sigma_i} \right)^2,
  \label{eq:chi2}
\end{equation}
where $y_i$ is the measured signal of channel $i$, $\hat{y}_i$ is its prediction with the vessel currents fitted to the channels in use, and $n$ is the number of channels in use.

\subsection{Null space and sensor loss}
\label{sec:nullspace}

The observations determine only part of $\jvec$. The vector $\jvec$ holds one current value per grid cell, 4089 in all. The matrix $\Hmat$ is $139 \times 4089$, so the sensors respond to at most 139 independent current modes, the right singular vectors of the whitened $\Hmat$. Fewer modes are resolved above the noise, 75 with all 124 channels in use (appendix~\ref{app:operator}). We split $\jvec = \jvec_\parallel + \jvec_\perp$, where $\jvec_\parallel$ lies in the row space of $\Hmat$, which the observations see, and $\jvec_\perp$ lies in its null space, which they do not see. The answer of a reconstruction is therefore set by what fills $\jvec_\perp$, that is, by the prior.

The prior of LIUQE combines force balance with a parametrized current profile~\cite{Moret2015}. LIUQE models the current density in the form $\jphi = R\,p'(\psi_\mathrm{N}) + FF'(\psi_\mathrm{N})/(\mu_0 R)$, with $p'$ and $FF'$ usually taken as low-order polynomials~\cite{Moret2015}. Here $\psi_\mathrm{N}$ is the normalized poloidal flux, $p'$ is the derivative of the pressure $p$ with respect to the poloidal flux, and $FF'$ is the product of the poloidal current function $F = RB_\phi$ and its derivative. Moret et al~\cite{Moret2015} write $F$ as $T$ and measure the flux in webers, which adds a factor $2\pi$ to the expression. This form follows from the ideal magnetohydrodynamic force balance $\nabla p = \bm{J} \times \bm{B}$~\cite{Moret2015}, where $\bm{J}$ is the current density and $\bm{B}$ is the magnetic field. The degrees of freedom are a few polynomial coefficients, so these coefficients determine $\jvec_\perp$.

Losing a channel is equivalent to deleting its row from $\Hmat$. With a sensor mask $\Smat$, the observations become $\Smat\yvec = \Smat\Hmat\jvec + \Smat\Hmat_\mathrm{a}\Ivec_\mathrm{a} + \Smat\Hmat_\mathrm{v}\Ivec_\mathrm{v}$. Fewer rows shrink the row space, and the null space grows by the same amount. For one random mask, the 75 resolved modes of the full sensor set drop to 37 at dropout 0.5 and to 10 at dropout 0.9.

\begin{figure}[t]
  \centering
  \includegraphics{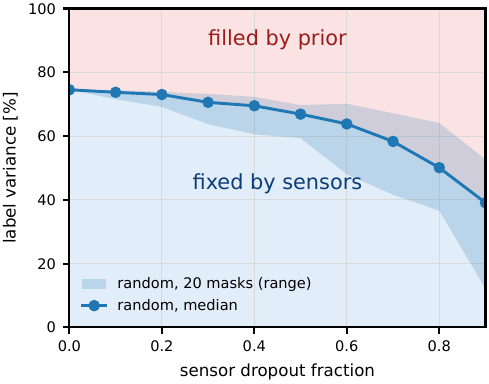}
  \caption{Fraction of the variance of the label $\jphi$ fixed by the magnetic sensors against the dropout fraction, with the vessel currents marginalized.}
  \label{fig:variance}
\end{figure}

Figure~\ref{fig:variance} measures how much of the label variance the sensors fix. We treat the label $\jphi$ as Gaussian. Twenty scalars are measured without the magnetic sensors and therefore remain under sensor loss. They are the 18 PF coil currents, the plasma current $I_\mathrm{p}$ and the product $RB_\mathrm{t}$ of the major radius and the toroidal magnetic field. The diffusion prior of section~\ref{sec:prior} takes them as input, so the sensors and the prior only need to fix the part of $\jphi$ that these scalars do not explain. We regressed the label $\jphi$ on the 20 scalars and take $\mathsf{K}$ as the covariance of the residual. We computed $\mathsf{K}$ from \num{35026} frames, one in every 10 converged frames of 799 of the training shots. Conditioning on the masked observations gives the posterior covariance
\begin{equation}
  \mathsf{K}_\mathrm{post} = \left( \mathsf{K}^{-1} + \Hmat_\mathrm{w}^\top \Hmat_\mathrm{w} \right)^{-1},
  \label{eq:kpost}
\end{equation}
where $\Hmat_\mathrm{w} = \Wmat_\mathrm{v}^{1/2}\Smat\Sigmat^{-1/2}\Hmat$ is the masked and whitened response with the vessel currents eliminated as in the sampler by the matrix $\Wmat_\mathrm{v}^{1/2}$ of equation~\eqref{eq:wv}. The sensors fix the fraction $1 - \mathrm{tr}\,\mathsf{K}_\mathrm{post} / \mathrm{tr}\,\mathsf{K}$ of the variance, and the rest is left to the prior. The full sensor set fixes 74.5\% and leaves 25.5\% to the prior. We drew 20 random masks for each dropout fraction. At dropout 0.9, the remaining sensors fix a median of 39.1\%, and 12.2--52.6\% depending on the mask.

After sensor loss, LIUQE fills the enlarged share with force balance and polynomial profiles. The diffusion prior fills it with its training distribution. The question of this paper is how close to the label each of the two priors comes when it fills this share.

\section{Method}
\label{sec:method}

\subsection{Diffusion prior}
\label{sec:prior}

The diffusion model generates a single channel, the toroidal current density $\jphi$ on the $87 \times 47$ grid of the observation model. It is trained on the labels of the training shots. We follow the formulation of Karras et al~\cite{Karras2022}, called EDM. A diffusion model is trained to remove noise. During training, Gaussian noise of a random level is added to a clean field, and a neural network learns to recover the clean field. Sampling starts from pure noise and removes it step by step as the noise level decreases, so that the result is a field from the training distribution. The model works in the normalized coordinates $\xvec = (\jvec - \bar{j}\,\onevec)/s$, where $\bar{j}$ and $s$ are the mean and the standard deviation of the label values over all grid points of the training frames, and $\onevec$ is the vector of ones. A noisy field is $\xvec_t = \xvec_0 + \sigma_t\bm{n}$, where $\xvec_0$ is a clean field and $\bm{n}$ is standard Gaussian noise. The noise level $\sigma_t$ is thus the standard deviation of the noise in units of $s$.

The denoiser also takes as input the 20 scalars of section~\ref{sec:nullspace}, which remain under sensor loss. The model therefore learns the distribution of $\jphi$ given these scalars. Giving such an input to a diffusion model is called conditioning, and we call the 20 scalars the conditioning vector $\bm{c}$. An embedding of the vector is added to the embedding of the noise level $\sigma_t$, and both enter every block of the denoiser.

The denoiser $D_\theta(\xvec_t; \sigma_t, \bm{c})$ is a neural network with parameters $\theta$. It is trained to return the clean field by minimizing
\begin{equation}
  \mathcal{L}(\theta) = \mathbb{E}_{\xvec_0, \bm{c}, \sigma, \bm{n}} \left[ \lambda(\sigma) \left\| D_\theta(\xvec_0 + \sigma\bm{n}; \sigma, \bm{c}) - \xvec_0 \right\|^2 \right],
  \label{eq:edm-loss}
\end{equation}
where $\xvec_0$ and $\bm{c}$ come from the same training frame, $\bm{n}$ is standard Gaussian noise, $\ln\sigma$ is drawn from $\mathcal{N}(P_\mathrm{mean}, P_\mathrm{std}^2)$, and $\lambda(\sigma) = (\sigma^2 + 1)/\sigma^2$ is the weight of EDM. The minimizer of this loss is the conditional mean of the clean field, so the trained denoiser gives $D_\theta(\xvec_t; \sigma_t, \bm{c}) \approx \mathbb{E}[\xvec_0 \mid \xvec_t, \bm{c}]$. It also gives the score of the noisy distribution~\cite{Karras2022},
\begin{equation}
  \nabla_{\xvec_t} \log p(\xvec_t \mid \bm{c}) \approx \frac{D_\theta(\xvec_t; \sigma_t, \bm{c}) - \xvec_t}{\sigma_t^2}.
  \label{eq:tweedie}
\end{equation}
The sampler of section~\ref{sec:sampling} uses the denoiser output as its clean estimate. From EDM we also take the scaling of the denoiser input and output with $\sigma_t$ and the decreasing sequence of noise levels used in sampling. The denoiser is a UNet~\cite{Ronneberger2015}. Table~\ref{tab:prior-settings} in appendix~\ref{app:settings} lists its architecture and the training settings, including $P_\mathrm{mean}$ and $P_\mathrm{std}$.

\subsection{Forward operator}
\label{sec:forward}

The matrices $\Hmat$, $\Hmat_\mathrm{a}$ and $\Hmat_\mathrm{v}$ of equation~\eqref{eq:obs} are the sensor response matrices of LIUQE in the open-source MEQ (MATLAB equilibrium) toolbox~\cite{Moret2015,Carpanese2021}\footnote{\url{https://gitlab.epfl.ch/spc/public/meq/meq}}, used without change. The diffusion reconstruction and the LIUQE baseline therefore share these matrices. We compute $\chi^2$ for both methods with the same $\sigma_i$, and with the vessel currents fitted in the same way by equation~\eqref{eq:vessel-fit} below. The matrices have two kinds of rows. A flux loop row gives the poloidal flux at the loop, and a probe row gives the magnetic field along the probe direction. Appendix~\ref{app:operator} gives the block structure of the matrices and the expressions of their elements.

The PF coil currents $\Ivec_\mathrm{a}$ are measured, so we move their contribution to the data side of equation~\eqref{eq:obs}. The data become $\yvec - \Hmat_\mathrm{a}\Ivec_\mathrm{a}$. The 104 vessel filament currents $\Ivec_\mathrm{v}$ are not measured, so we marginalize them with the Gaussian prior $\Ivec_\mathrm{v} \sim \mathcal{N}(\bm{0}, \sigma_\mathrm{v}^2 \Imat)$. The marginalized likelihood is still Gaussian. Its covariance grows from the noise covariance $\Sigmat = \mathrm{diag}(\sigma_i^2)$ to $\Sigmat + \sigma_\mathrm{v}^2 \Hmat_\mathrm{v}\Hmat_\mathrm{v}^\top$. In whitened coordinates, with the masked vessel response $\tilde{\Hmat}_\mathrm{v} = \Smat\Sigmat^{-1/2}\Hmat_\mathrm{v}$, this covariance is $\Imat + \sigma_\mathrm{v}^2 \tilde{\Hmat}_\mathrm{v}\tilde{\Hmat}_\mathrm{v}^\top$. Its inverse square root is
\begin{equation}
  \Wmat_\mathrm{v}^{1/2} = \left( \Imat + \sigma_\mathrm{v}^2 \tilde{\Hmat}_\mathrm{v}\tilde{\Hmat}_\mathrm{v}^\top \right)^{-1/2}.
  \label{eq:wv}
\end{equation}
Multiplying the whitened rows and data by $\Wmat_\mathrm{v}^{1/2}$ restores unit noise covariance, so $\Ivec_\mathrm{v}$ no longer appears in the likelihood. The order is the mask, then $\Sigmat^{-1/2}$, then $\Wmat_\mathrm{v}^{1/2}$. Because $\tilde{\Hmat}_\mathrm{v}$ contains the mask, $\Wmat_\mathrm{v}^{1/2}$ is rebuilt for each mask. Appendix~\ref{app:operator} shows which directions of the observations it suppresses. We set $\sigma_\mathrm{v} = \qty{1}{\kilo\ampere}$, the value that LIUQE uses to regularize the vessel currents when it computes the labels.

The plasma current $I_\mathrm{p}$ and the limiter condition also enter as rows, which we call constraint rows. The plasma current gives one row, $I_\mathrm{p} = \Delta A\,\onevec^\top \jvec$, where $\Delta A$ is the area of a grid cell. This row sets the total current on the grid to the measured plasma current. The condition $\jphi = 0$ outside the limiter gives one row for each grid point outside the limiter. Like a sensor row, each constraint row is divided by its standard deviation. Table~\ref{tab:sampler-settings} lists these standard deviations. The mask $\Smat$ deletes sensor rows only. The $I_\mathrm{p}$ row and the limiter rows always remain.

\subsection{Sampling}
\label{sec:sampling}

We sample the posterior with DAPS~\cite{Zhang2024}, which runs through the decreasing noise levels $\sigma_t$ of the diffusion model. The sampler starts from Gaussian noise at the largest noise level. At each noise level, DAPS repeats three steps, numbered 1 to 3 in figure~\ref{fig:method}. The denoiser maps the noisy state $\jvec_t$ to the clean estimate
\begin{equation}
  \hat{\jvec}_0 = \bar{j}\,\onevec + s\,D_\theta(\xvec_t; \sigma_t, \bm{c}), \qquad \xvec_t = \frac{\jvec_t - \bar{j}\,\onevec}{s},
  \label{eq:denoise-j}
\end{equation}
which is the conditional mean of section~\ref{sec:prior} in physical units. Because the noise level $\sigma_t$ is in units of $s$, the noise on $\jvec_t$ has standard deviation $s\sigma_t$. The correction moves $\hat{\jvec}_0$ toward the data and gives the corrected estimate $\jvec_0^\ast$. The renoising adds Gaussian noise of the next noise level $\sigma_{t'}$ and gives the next noisy state $\jvec_{t'} = \jvec_0^\ast + s\,\sigma_{t'}\bm{n}$. After the last step, $\jvec_0^\ast$ is the sample. The right part of figure~\ref{fig:method} shows how the flux is derived from the mean of the samples, which section~\ref{sec:derived-psi} describes. The correction acts on the clean estimate $\hat{\jvec}_0$, so it needs no differentiation through the denoiser. In DAPS, the correction draws a sample by Langevin dynamics, an iterative sampler driven by the gradient of the log-posterior. Our observations are linear, so the objective of the correction is quadratic and has a closed-form minimizer. We replace the Langevin sampling with this minimizer. The correction is then deterministic, and the samples of a frame differ only through the initial noise and the renoising.

The correction step minimizes a quadratic objective,
\begin{equation}
  \jvec_0^\ast = \arg\min_{\jvec} \, \|\Amat\jvec - \bvec\|^2 + \lambda \|\jvec - \hat{\jvec}_0\|^2,
  \qquad
  \lambda = \frac{1}{\zeta s^2 (\sigma_t^2 + \epsilon)},
  \label{eq:correction}
\end{equation}
with $\epsilon = 10^{-4}$. Here $\zeta$ is a data weight, $\epsilon$ keeps $\lambda$ bounded at the smallest noise levels, and the operator $\Amat$ and the data $\bvec$ stack the masked sensor rows on the constraint rows,
\begin{equation}
  \Amat = \begin{bmatrix} \Hmat_\mathrm{w} \\ \Amat_\mathrm{c} \end{bmatrix},
  \qquad
  \bvec = \begin{bmatrix} \Wmat_\mathrm{v}^{1/2}\Smat\Sigmat^{-1/2}(\yvec - \Hmat_\mathrm{a}\Ivec_\mathrm{a}) \\ \bvec_\mathrm{c} \end{bmatrix}.
  \label{eq:stack}
\end{equation}
The sensor block $\Hmat_\mathrm{w} = \Wmat_\mathrm{v}^{1/2}\Smat\Sigmat^{-1/2}\Hmat$ is the masked and whitened plasma response of section~\ref{sec:nullspace}, in which $\Wmat_\mathrm{v}^{1/2}$ eliminates the vessel currents. The rows of $\Amat_\mathrm{c}$ and the entries of $\bvec_\mathrm{c}$ are the constraint rows of section~\ref{sec:forward} and their values, each divided by its standard deviation. Up to a factor of 2 and an additive constant, the first term of equation~\eqref{eq:correction} is the negative log-likelihood of the whitened observations and constraint rows, and the second term is the negative log-density of a Gaussian centered on $\hat{\jvec}_0$ with variance $\zeta s^2 (\sigma_t^2 + \epsilon)$. Here $\sigma_t$ is the noise level of the present step, and $s$ converts it to the units of $\jvec$. As $\sigma_t$ decreases, $\lambda$ grows and $\jvec_0^\ast$ stays closer to $\hat{\jvec}_0$. A larger $\zeta$ pulls the solution further toward the data. The objective is quadratic, so the correction has the closed-form solution $\jvec_0^\ast = \hat{\jvec}_0 + (\Amat^\top\Amat + \lambda\Imat)^{-1}\Amat^\top(\bvec - \Amat\hat{\jvec}_0)$. The matrix $\Amat^\top\Amat$ does not depend on the frame, so we factorize it once for each mask. The change $\jvec_0^\ast - \hat{\jvec}_0$ lies in the row space of $\Amat$, so the correction changes only the components that the observations see and keeps the other components of $\hat{\jvec}_0$. As $\lambda \to 0$, and when $\Amat$ has full row rank, it becomes the orthogonal projection of $\hat{\jvec}_0$ onto the solutions of $\Amat\jvec = \bvec$. The correction is thus a regularized projection of the clean estimate, whereas projection methods~\cite{Song2022} replace the observed component of the noisy sample.

We run the sampler $K$ times per frame with independent noise and obtain $K$ samples. A sample is excluded if its $I_\mathrm{p}$ error, the relative difference between $\Delta A\,\onevec^\top\jvec$ and the measured $I_\mathrm{p}$, is large compared with the other samples of the frame. The reported reconstruction is the mean of the remaining samples. Table~\ref{tab:sampler-settings} gives $K$ and the exclusion threshold. The $K$ samples are not exact samples of the posterior. Section~\ref{sec:uncertainty} tests whether their spread can still serve as an uncertainty. The parameter $\zeta$ sets the relative weight of the data term. We choose $\zeta$ on the validation shots with all sensors in use. As we increase $\zeta$, we take the first value at which the $\chi^2$ of the reconstruction falls below the $\chi^2$ of the label. The evaluation shots are not used for this choice. The same $\zeta$ is used for every dropout setting.

Appendix~\ref{app:settings} gives the correction in normalized coordinates, the factorization, the number of steps and the $\zeta$ scan. Figure~\ref{fig:sawtooth} follows the label distance of $\jphi$ along the sampling steps. Section~\ref{sec:likelihood} uses it to examine what the correction leaves in the final solution.

\subsection{Derived poloidal flux}
\label{sec:derived-psi}

The poloidal flux follows from the reconstructed current $\jvec$, the mean of the retained samples, through equation~\eqref{eq:psi} with the vessel currents replaced by an estimate $\hat{\Ivec}_\mathrm{v}$,
\begin{equation}
  \psivec = \Gmat\jvec + \Gmat_\mathrm{a}\Ivec_\mathrm{a} + \Gmat_\mathrm{v}\hat{\Ivec}_\mathrm{v}.
\end{equation}
The flux matrices $\Gmat$, $\Gmat_\mathrm{a}$ and $\Gmat_\mathrm{v}$ also come from LIUQE in the MEQ toolbox, like the sensor response matrices of section~\ref{sec:forward}. The estimate $\hat{\Ivec}_\mathrm{v}$ is a regularized least-squares estimate. We fix the reconstructed $\jvec$ and fit the vessel currents to the sensors that remain after the mask $\Smat$,
\begin{equation}
  \hat{\Ivec}_\mathrm{v} = \arg\min_{\Ivec_\mathrm{v}} \, \big\|\Smat\Sigmat^{-1/2}\big(\yvec - \Hmat\jvec - \Hmat_\mathrm{a}\Ivec_\mathrm{a} - \Hmat_\mathrm{v}\Ivec_\mathrm{v}\big)\big\|^2 + \frac{\|\Ivec_\mathrm{v}\|^2}{\sigma_\mathrm{v}^2}.
  \label{eq:vessel-fit}
\end{equation}
The factor $\Sigmat^{-1/2}$ whitens the residual, and $\Smat$ keeps only the remaining channels. The standard deviation $\sigma_\mathrm{v}$ is the same one that section~\ref{sec:forward} uses to marginalize the vessel currents. \emph{We impose no force balance and no profile constraint on $\psivec$.} The flux is a consequence of $\jvec$.

\subsection{Data and evaluation}
\label{sec:data-eval}

The data are the flattop LIUQE labels of 933 KSTAR shots, \#38843--\#42391, \num{403450} frames in total. They include negative-triangularity discharges. Of these, 803 shots (\num{350253} frames) are used for training, 27 validation shots for the choice of $\zeta$, and 37 shots for evaluation. The evaluation shots form three blocks of consecutive shot numbers. The eight shots on either side of each block are excluded from training.

The labels are full-sensor reconstructions computed with LIUQE in the MEQ toolbox~\cite{Moret2015,Carpanese2021}. Table~\ref{tab:label-settings} in appendix~\ref{app:settings} lists the LIUQE settings, among them the profile basis and the weights of the measurements. Frames on which LIUQE did not converge were excluded from the labels.

We simulate sensor loss by random dropout. Random dropout switches off a fixed fraction of the 124 channels in use, chosen at random. We use the dropout fractions 0.0, 0.1, \ldots, 0.9, which gives ten settings. The mask $\Smat$ of each setting is drawn once and applied to all evaluation frames. The masks of different settings were drawn independently and are not nested.

The baseline is LIUQE from MEQ, which we can rerun with any sensor mask. LIUQE also produced the labels. Like EFIT~\cite{Lao1985}, it parametrizes the current density with a linear combination of base functions~\cite{Moret2015}. We run LIUQE again with the same mask as the diffusion reconstruction, that is, with zero weight on the removed channels. In its objective, LIUQE weights the channels with its own error estimates, which differ from the $\sigma_i$. The $\chi^2$ of LIUQE is therefore computed with weights that differ from those it minimizes. LIUQE solves the frames of a shot in time order and starts each frame from the solution of the previous frame. The labels were computed in the same way, and we keep this mode of operation. The previous frame is also solved with the same mask, so the evaluation models sensor loss that persists throughout the shot. We do not treat sudden sensor failures within a shot. The errors of frames on which LIUQE does not converge are included in the results. For each setting, 1--61 of the 370 evaluation frames defined below did not converge. As a sensitivity check, section~\ref{sec:results} also reports a version in which both methods are restricted to the frames on which LIUQE converged.

The label distance is the relative $L_2$ distance of a reconstruction from the label. For $\jphi$ and for $\psivec$ it is
\begin{equation}
  \frac{\|\jvec - \jvec^\mathrm{label}\|}{\|\jvec^\mathrm{label}\|}
  \qquad \text{and} \qquad
  \frac{\|\psivec - \psivec^\mathrm{label}\|}{\|\psivec^\mathrm{label}\|},
  \label{eq:label-distance}
\end{equation}
expressed in percent, where $\|\cdot\|$ is the Euclidean norm over the grid points. For the diffusion reconstruction $\psivec$ is derived from $\jvec$, and for LIUQE it is the flux of LIUQE itself. From each evaluation shot, we take 10 flattop label frames at equal spacing in the time-ordered list of label frames, including the first and the last. This gives the 370 evaluation frames. The label distance measures agreement with the full-sensor LIUQE reconstruction. It is not a distance from the actual equilibrium of the plasma.

Because the label is itself a LIUQE reconstruction, this distance treats the two methods asymmetrically. At dropout 0.0, the label distance of LIUQE is 0 by definition. That of the diffusion reconstruction is the distance between two full-sensor reconstructions that fill the null-space component $\jvec_\perp$ with different priors, not an error caused by sensor loss.

The fit of both methods is the $\chi^2$ of equation~\eqref{eq:chi2} over the channels that remain after the mask. For both methods, the prediction $\hat{y}_i$ is computed from the reconstructed $\jvec$ with the vessel currents $\hat{\Ivec}_\mathrm{v}$ of equation~\eqref{eq:vessel-fit}. Its reference is the full-sensor $\chi^2$ of the labels on the validation shots, 0.675, taken as the median over shots of the per-shot medians. The $\sigma_i$ were calibrated from the residuals of the labels, so we use the value of $\chi^2$ to compare methods and not as an absolute measure of fit.

\section{Results}
\label{sec:results}

\begin{figure}[tbp]
  \centering
  \includegraphics{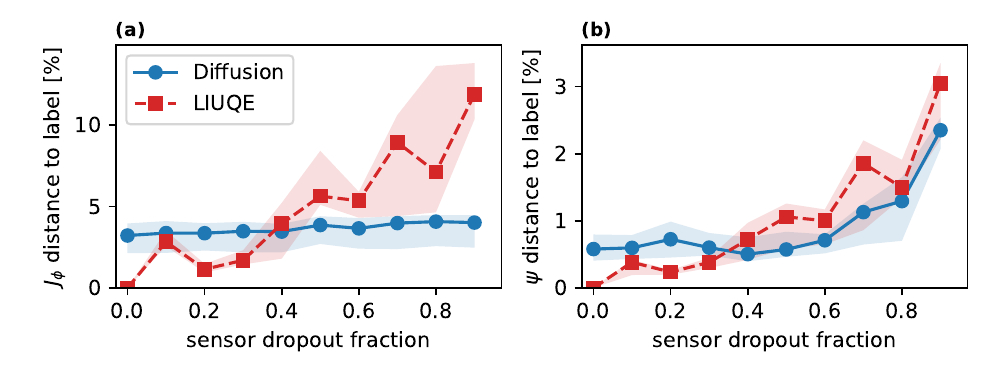}
  \caption{Label distance of (a)~$\jphi$ and (b)~$\psivec$ against the dropout fraction. Lines show the median over the 37 evaluation shots, and bands span the 25--75th percentiles.}
  \label{fig:dropout}
\end{figure}

Figure~\ref{fig:dropout}(a) shows the label distance of $\jphi$ under random dropout on the 37 KSTAR evaluation shots. The horizontal axis is the dropout fraction of the ten settings. For each shot, we take the median label distance over its 10 evaluation frames. Each line is the median of these per-shot medians over the 37 shots, and the band spans their 25--75th percentiles. A lower value means closer agreement with the label. The label distance of the diffusion reconstruction is flat over the ten settings, at 3.22--4.07\%. The label distance of LIUQE rises as the dropout fraction grows and reaches 11.89\% at dropout 0.9.

We then compare the two methods shot by shot. For each setting, we count the shots on which the per-shot median label distance of the diffusion reconstruction is lower than that of LIUQE. At dropout 0.5--0.9, the label distance of the diffusion reconstruction is lower than that of LIUQE on 27--37 of the 37 shots. At dropout 0.9, it is lower on all 37 shots. At dropout 0.2 and 0.3, it is lower on 1 and 2 of the 37 shots, respectively. The median curve of LIUQE drops from 2.84\% at dropout 0.1 to 1.15\% at dropout 0.2. The masks of the settings are drawn independently and are not nested, so the curve also depends on which channels were removed.

The two methods fit the remaining channels about equally well. Over the ten settings, the median $\chi^2$ is 0.58--0.70 for the diffusion reconstruction and 0.50--0.75 for LIUQE. Both are close to the reference value 0.675, the full-sensor $\chi^2$ of the labels on the validation shots.

Figure~\ref{fig:dropout}(b) shows the label distance of the poloidal flux $\psivec$ in the same layout as panel (a). At dropout 0.5--0.9, the label distance of the diffusion reconstruction is lower than that of LIUQE on 21--36 of the 37 shots. At dropout 0.1--0.3, LIUQE is lower on 36--37 of the 37 shots. At dropout 0.9, the label distance of $\psivec$ is 2.35\% for the diffusion reconstruction and 3.05\% for LIUQE. As a sensitivity check, we keep only the frames on which LIUQE converged. The diffusion reconstruction is lower at dropout 0.5--0.9 on 27--37 shots for $\jphi$ and on 21--29 shots for $\psivec$, and LIUQE remains lower for $\psivec$ at dropout 0.1--0.3.

The label distances of $\psivec$ and $\jphi$ should not be compared in magnitude, because the coil term $\Gmat_\mathrm{a}\Ivec_\mathrm{a}$ dominates the norm of the label flux and $\Gmat$, which sums over all grid currents, compresses the shape error of $\jphi$.

The evaluation shots form three blocks of lower single null (LSN, 15 shots), upper single null (USN, 7 shots) and double null (DN, 15 shots) discharges. We group the frames into these three shape classes by their block and by the topology of the label $\psivec$, as appendix~\ref{app:shapes} describes. Appendix~\ref{app:shapes} also maps the errors of $\jphi$ and $\psivec$ over the grid for each class.

\begin{figure}[p]
  \centering
  \includegraphics{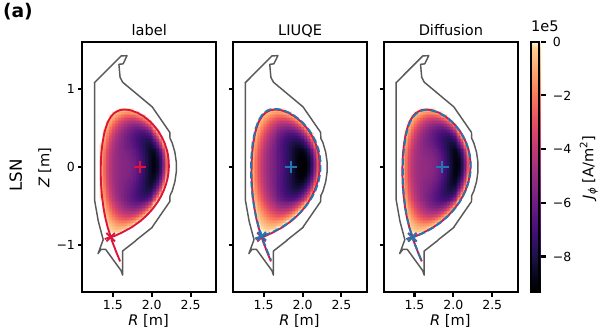}\par\medskip
  \includegraphics{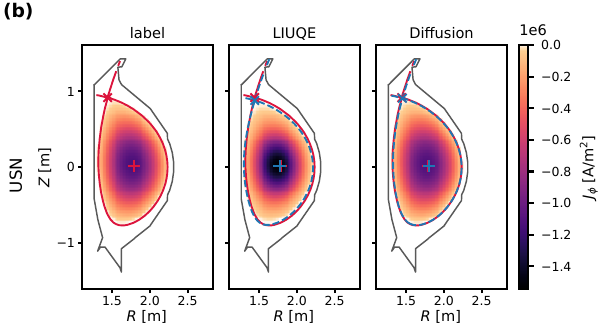}\par\medskip
  \includegraphics{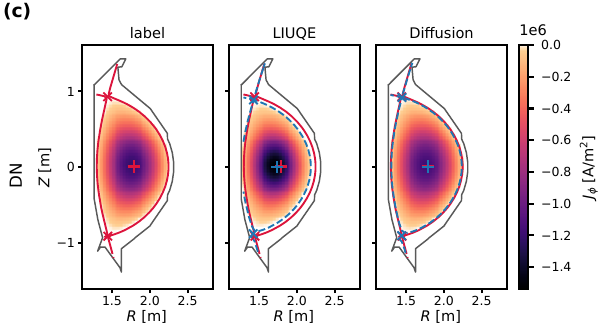}
  \caption{Label, LIUQE and diffusion reconstructions of $\jphi$ at dropout 0.9 for the (a)~LSN, (b)~USN and (c)~DN classes. Each row is the frame with the median label distance of the diffusion reconstruction in its class. Red solid and blue dashed lines are the separatrices of the label and the reconstruction, markers in the same colors are the magnetic axis and X-points, and the gray line is the limiter.}
  \label{fig:shapes}
\end{figure}

Table~\ref{tab:shapes} lists the label distance by shape class at dropout 0.5 and 0.9. At both dropout fractions, the diffusion reconstruction has the lower median label distance of $\jphi$ and of $\psivec$ in all three shape classes. The difference is smallest for $\psivec$ of the DN class at dropout 0.9, 1.77\% against 1.99\%.

\begin{table}[tbp]
  \centering
  \caption{Median label distance [\%] by shape class at dropout 0.5 and 0.9. Frames on which LIUQE did not converge are included.}
   \label{tab:shapes}
  \footnotesize
  \begin{tabular}{@{}llcccc@{}}
    \toprule
    & & \multicolumn{2}{c}{$\jphi$ label distance [\%]} & \multicolumn{2}{c}{$\psivec$ label distance [\%]} \\
    \cmidrule(lr){3-4} \cmidrule(lr){5-6}
    Shape & Dropout & Diffusion & LIUQE & Diffusion & LIUQE \\
    \midrule
    LSN & 0.5 & 4.24 & 8.62 & 0.82 & 1.28 \\
        & 0.9 & 4.33 & 11.43 & 2.48 & 3.35 \\
    USN & 0.5 & 2.14 & 5.21 & 0.55 & 1.00 \\
        & 0.9 & 2.32 & 10.11 & 2.35 & 2.56 \\
    DN  & 0.5 & 3.22 & 5.01 & 0.44 & 0.56 \\
        & 0.9 & 4.01 & 13.84 & 1.77 & 1.99 \\
    \bottomrule
  \end{tabular}
\end{table}

Figure~\ref{fig:shapes} shows one frame of each shape class at dropout 0.9. In each class, it is the frame with the median label distance of the diffusion reconstruction. On these frames, the label distance of $\jphi$ is 4.07\% against 8.57\% for LSN, 2.42\% against 21.53\% for USN and 2.58\% against 24.01\% for DN (diffusion reconstruction against LIUQE). Because the frames are chosen by the diffusion reconstruction, the LIUQE panels of USN and DN are worse than the class medians of table~\ref{tab:shapes}.

\FloatBarrier
\section{Discussion}
\label{sec:discussion}

The diffusion reconstruction uses physics differently from LIUQE and from neural network reconstructions that put the Grad--Shafranov equation into the training loss~\cite{Joung2023gs,Rossi2023}. LIUQE builds force balance into its prior by writing $\jphi$ in terms of the profiles $p'$ and $FF'$. In the diffusion reconstruction, the prior learned from the labels sets the shape of $\jphi$, and physics enters only through the observation model of equation~\eqref{eq:obs}, which is linear in $\jvec$. Bayesian current tomography~\cite{Svensson2008} and decoupled diffusion solvers~\cite{Lin2026} split the problem in the same way. They put a prior on the unknown current or field and let the physics enter only through the forward operator of the likelihood.

\subsection{Likelihood correction}
\label{sec:likelihood}

\begin{figure}[tbp]
  \centering
  \includegraphics{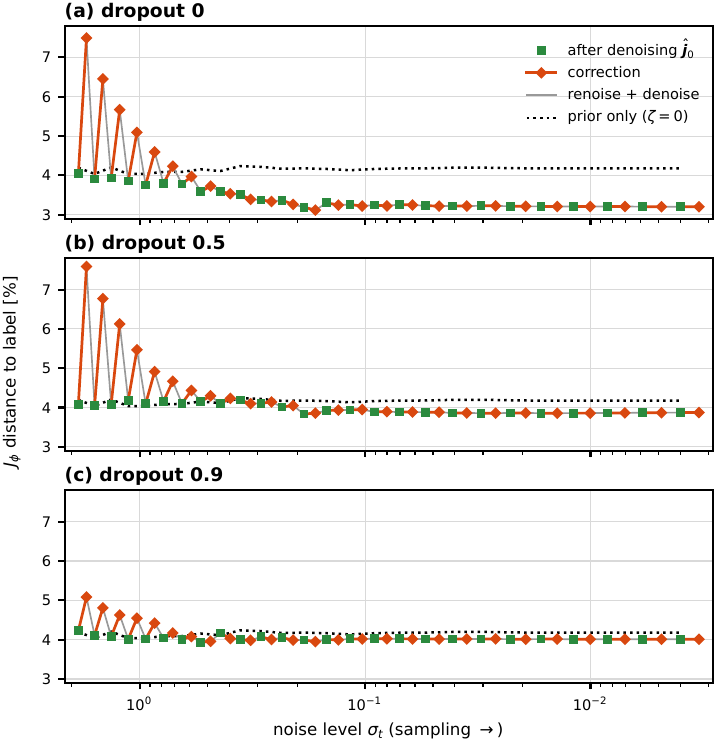}
  \caption{Label distance of the mean of the $K$ samples of $\jphi$ along the last sampling steps, $\sigma_t \le 2$, at dropout (a)~0, (b)~0.5 and (c)~0.9.}
  \label{fig:sawtooth}
\end{figure}

Section~\ref{sec:nullspace} showed that the remaining sensors fix a smaller fraction of the variance of $\jphi$ as the dropout grows. The conditional prior therefore determines more of the reconstruction, and its conditioning vector does not depend on the dropout. The null-space component $\jvec_\perp$ comes only from the prior, and Bhadra et al~\cite{Bhadra2021} call errors in this component hallucinations of the prior. To separate the two contributions, we compare the reconstruction with a prior-only reconstruction. It uses the same sampler, random noise and conditioning vector, with the correction switched off ($\zeta = 0$). This is the limit $\lambda \to \infty$ of equation~\eqref{eq:correction}, in which $\jvec_0^\ast = \hat{\jvec}_0$, so the prior-only reconstruction is the same for every dropout setting. Its label distance of $\jphi$ is 4.18\%, against 3.22\%, 3.86\% and 4.01\% for the reconstruction at dropout 0, 0.5 and 0.9. The prior alone thus comes close to the label, as the dotted line of figure~\ref{fig:sawtooth} shows along the whole sampling. Paired shot by shot, the reconstruction is lower on 37, 24 and 29 of the 37 shots, and the median over shots of the difference is $-0.58$, $-0.13$ and $-0.12$ percentage points.

Figure~\ref{fig:sawtooth} shows when the correction acts. At $\sigma_t = 1.87$, the correction moves the estimate away from the label, to 7.49\%, 7.59\% and 5.08\% at dropout 0, 0.5 and 0.9, and the renoising of the next step erases this change. The same holds for every correction at $\sigma_t > 1$. In linear inverse problems, the measurement term of the posterior score is also weak relative to the prior score at high noise~\cite{Meng2026}. The corrections at $\sigma_t \le 1$ survive. At dropout 0, the denoiser output after each of them lies below the prior-only curve. At dropout 0.5 and 0.9, it stays within 0.1 percentage points of the prior-only curve down to $\sigma_t \approx 0.3$ and $0.5$, respectively, and separates from it after that.

The correction pays off where the prior disagrees with the sensors. We rank the 370 evaluation frames by how much the $\chi^2$ of the prior-only reconstruction, over all 124 channels, exceeds that of the label. On the top 10\% of this ranking, the reconstruction at dropout 0 lowers the median label distance of $\jphi$ from 9.09\% to 6.14\%. On the other frames, it lowers it from 3.59\% to 2.91\%. The gain of the correction is therefore largest on the frames where the prior alone fits the sensors worst.

\subsection{Vessel current estimation}
\label{sec:vessel}

The two methods also treat the vessel currents differently. LIUQE fits the vessel currents together with $\jphi$~\cite{Moret2015}. The diffusion reconstruction integrates $\Ivec_\mathrm{v}$ out of the likelihood when it samples $\jphi$. Afterwards, it fits $\hat{\Ivec}_\mathrm{v}$ to the remaining channels with $\jvec$ fixed and uses it only to derive $\psivec$. This fit is why the label distance of $\psivec$ rises at high dropout in figure~\ref{fig:dropout}(b) while that of $\jphi$ stays flat. We recomputed $\psivec$ from the diffusion $\jphi$ with $\hat{\Ivec}_\mathrm{v}$ fitted to all 124 channels instead, and its label distance became flat at 0.58--0.75\% over the ten settings. Conversely, with the label $\jphi$ and $\hat{\Ivec}_\mathrm{v}$ fitted to the remaining channels, the label distance of $\psivec$ at dropout 0.9 is 2.40\%, at the level of the 2.35\% of the reconstruction. The error of $\psivec$ at high dropout therefore comes from estimating the 104 vessel filament currents from a few sensors, not from the prior of $\jphi$.

\subsection{Uncertainty}
\label{sec:uncertainty}

Bayesian equilibrium inference gives a posterior distribution over current distributions~\cite{Svensson2008,vonNessi2013}, and the $K$ samples of the diffusion reconstruction invite the same use. Their spread, however, is not a calibrated uncertainty. Over the ten settings, the nominal 95\% interval built from the samples contains the label at only 43--48\% of the pixels. Appendix~\ref{app:spread} gives the definitions and further statistics. The samples differ only through the initial noise and the renoising. Their spread therefore does not contain a mismatch between the prior and the actual distribution, nor the error of the forward model. The underestimate grows away from the training distribution. With all sensors, the median of the standardized error $|z|$ is 1.9 times its calibrated value on the validation shots, which lie among the training shots. On the evaluation shots, whose nearest training shot is a median of 14 shot numbers away, it is 3.5 times. The sample spread should therefore not be used as an error bar of the reconstruction.

\section{Conclusion}
\label{sec:conclusion}

We reconstructed KSTAR equilibria under magnetic sensor dropout with a conditional diffusion prior over $\jphi$ alone. The physics enters only through a linear observation model built from the sensor response matrices of LIUQE, so the likelihood correction of the DAPS sampler has a closed-form solution. The poloidal flux $\psivec$ is derived from $\jphi$. On 37 KSTAR evaluation shots, the label distance of $\jphi$ stays at 3.22--4.07\% over ten random dropout settings, whereas that of LIUQE reaches 11.89\% at dropout 0.9. The prior carries most of the reconstruction, and its inputs, the PF coil currents, $I_\mathrm{p}$ and $RB_\mathrm{t}$, do not depend on the dropout. The remaining sensors add a correction that is largest where the prior fits them worst. The error of $\psivec$ still grows at high dropout, and this growth comes from the estimate of the vessel currents, not from the prior.

Improving the estimate of the vessel currents may reduce the error of $\psivec$ at high dropout. The candidates are a smaller prior standard deviation $\sigma_\mathrm{v}$ of the vessel currents, a few vessel eigenmodes and other constraints as in LIUQE, and we did not measure which of these works. The training distribution could be broadened with more discharges or with synthetic labels from a free-boundary equilibrium code, such as the forward solver FGS of the MEQ toolbox~\cite{Carpanese2021}. A shorter reconstruction time would allow between-shot analysis and control validation. The correction step runs on one CPU thread and takes about 80\% of the 1.1~s per frame, so it is the first target for a speed-up. Finally, the spread of the $K$ samples is not yet a calibrated uncertainty, because it does not contain the mismatch between the prior and the actual distribution.

\section*{Acknowledgements}
This work was supported by a National Research Foundation of Korea (NRF)
grant funded by the Korea government (MSIT) (Grant No.\ RS-2024-00346024).
During the preparation of this work, the authors used Claude Opus 5.5 (Anthropic)
to generate code and to translate the manuscript into English. After using this
tool, the authors reviewed and edited the content as needed and take full
responsibility for the content of the publication.

\clearpage
\appendix
\counterwithin{figure}{section}
\counterwithin{table}{section}
\counterwithin{equation}{section}
\renewcommand{\thefigure}{\thesection\arabic{figure}}
\renewcommand{\thetable}{\thesection\arabic{table}}
\section{Whitened observation operator}
\label{app:operator}

The response matrices of section~\ref{sec:forward} have the block form
\begin{equation}
  \begin{bmatrix} \Hmat & \Hmat_\mathrm{a} & \Hmat_\mathrm{v} \end{bmatrix}
  =
  \begin{blockarray}{cccl}
    \text{\footnotesize plasma} & \text{\footnotesize PF coils} & \text{\footnotesize vessel} & \\
    \begin{block}{[ccc]l}
      \Mmat_\mathrm{fp} & \Mmat_\mathrm{fa} & \Mmat_\mathrm{fv} & \text{\footnotesize FL} \\
      \Bmat_\mathrm{mp} & \Bmat_\mathrm{ma} & \Bmat_\mathrm{mv} & \text{\footnotesize MP, PLMP} \\
    \end{block}
  \end{blockarray}
\end{equation}
with 45 FL, 84 MP and 10 PLMP rows, and with 18 PF coils and 104 vessel filaments as current sources. Every block, and the flux matrix $\Gmat$, comes from one kernel given by the Biot--Savart law~\cite{Hutchinson2002,Moret2015}, the mutual inductance $M(\rvec, \rvec')$ between two coaxial circular filaments at positions $\rvec = (R, Z)$ and $\rvec'$ on the poloidal cross-section. It has an analytic expression in complete elliptic integrals of the first and second kind~\cite{Moret2015}. The grid cell area $\Delta A = \qty{1.40625e-3}{\metre\squared}$ converts the current density at grid point $k$, at position $\rvec_k$, into a filament current. For a flux loop $f$ at position $\rvec_f$,
\begin{equation}
  (\Mmat_\mathrm{fp})_{fk} = M(\rvec_f, \rvec_k)\,\Delta A .
\end{equation}
A magnetic probe $m$ at position $\rvec_m = (R_m, Z_m)$ measures the field along the angle $\theta_m$ to the $R$ axis. Projecting the field of a unit filament current, $B_R = -\partial_Z M / (2\pi R)$ and $B_Z = \partial_R M / (2\pi R)$ with derivatives taken with respect to the first argument~\cite{Moret2015}, onto this direction gives
\begin{equation}
  (\Bmat_\mathrm{mp})_{mk} = \frac{1}{2\pi R_m} \left( \sin\theta_m\,\partial_R - \cos\theta_m\,\partial_Z \right) M(\rvec_m, \rvec_k)\,\Delta A .
\end{equation}
The flux matrix on the grid is
\begin{equation}
  \Gmat_{ik} = \frac{M(\rvec_i, \rvec_k)\,\Delta A}{2\pi},
\end{equation}
where $i$ is a grid point. The flux loops measure in webers, and the division by $2\pi$ gives $\psi$ in \si{\weber\per\radian}.

The noise level differs from row to row. We multiply the observation rows by $\Sigmat^{-1/2}$, where $\Sigmat = \mathrm{diag}(\sigma_i^2)$ is the noise covariance of section~\ref{sec:obs-model}. After this step, the noise on every row has unit variance. With the mask $\Smat$ of section~\ref{sec:nullspace}, the whitened plasma response is $\Smat\Sigmat^{-1/2}\Hmat$. In these coordinates, the weighted least-squares objective is the squared Euclidean norm of the residual, and the $\chi^2$ of equation~\eqref{eq:chi2} is this squared norm divided by $n$.

Let $\gamma_i$ be the singular values of the whitened plasma response $\Smat\Sigmat^{-1/2}\Hmat$. A current perturbation of norm $s$ along the $i$th right singular vector changes the whitened observations by $\gamma_i s$, where $s$ is the standard deviation of the labels (section~\ref{sec:sampling}). This mode is therefore resolved only if $\gamma_i s > 1$. We counted these modes for one random mask at each dropout fraction. The full sensor set of 124 rows has 75 such modes. Dropout 0.5 leaves 37, and dropout 0.9 leaves 10. The fraction of the label variance that the sensors determine (figure~\ref{fig:variance}) falls with the dropout fraction in the same order as these mode counts.

Section~\ref{sec:forward} eliminates the vessel currents with the matrix $\Wmat_\mathrm{v}^{1/2}$ of equation~\eqref{eq:wv}. Let $\tilde{\Hmat}_\mathrm{v} = \bm{\mathsf{U}}\,\mathrm{diag}(\nu_i)\,\bm{\mathsf{V}}^\top$ be the singular value decomposition, with singular values $\nu_i$. Then
\begin{equation}
  \Wmat_\mathrm{v}^{1/2} = \Imat - \bm{\mathsf{U}}\,\mathrm{diag}\!\left( 1 - \frac{1}{\sqrt{1 + \sigma_\mathrm{v}^2\nu_i^2}} \right) \bm{\mathsf{U}}^\top .
\end{equation}
The operator scales the component along the $i$th left singular vector by $(1 + \sigma_\mathrm{v}^2\nu_i^2)^{-1/2}$ and leaves the orthogonal complement unchanged. For $\sigma_\mathrm{v}\nu_i \gg 1$ this factor approaches zero, so these directions are nearly removed. They are projected out only in the limit of large $\sigma_\mathrm{v}\nu_i$. For $\sigma_\mathrm{v}\nu_i \ll 1$ the factor approaches one, and these directions remain.

\section{Training and reconstruction settings}
\label{app:settings}

\begin{figure}[t]
  \centering
  \includegraphics{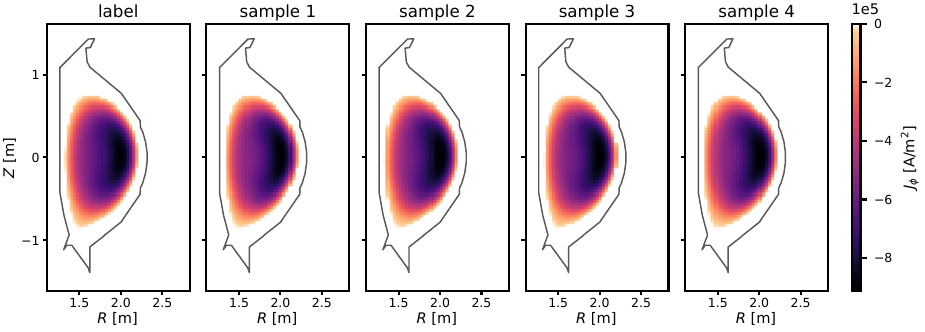}
  \caption{Label and four samples of the conditional diffusion prior for one frame, drawn without sensor data. White marks $\jphi = 0$.}
  \label{fig:prior}
\end{figure}

Table~\ref{tab:prior-settings} lists the architecture and the training settings of the denoiser of section~\ref{sec:prior}. The label standard deviation in the table is the scale $s$ of equation~\eqref{eq:correction}. During training, the conditioning vector was replaced by the zero vector with probability 0.1, as in classifier-free guidance~\cite{Ho2022}, but sampling uses the conditional denoiser only. Figure~\ref{fig:prior} shows the label and four samples of the prior for one frame, drawn under its conditioning vector without sensor data.

\begin{table}[t]
  \centering
  \caption{Settings of the conditional diffusion prior.}
  \label{tab:prior-settings}
  \small
  \begin{tabular}{@{}l>{\raggedright\arraybackslash}p{0.5\textwidth}@{}}
    \toprule
    Setting & Value \\
    \midrule
    Denoiser & UNet \\
    Encoder stages & 3, widths 64, 128, 256 \\
    Middle blocks & 2 \\
    Decoder stages & 3 \\
    Noise-level embedding & Fourier, size 128 \\
    Conditioning & 20 scalars, multilayer perceptron, added to the noise-level embedding \\
    Embedding input of each block & FiLM layer~\cite{Perez2018} \\
    Label mean & \qty{-89398}{\ampere\per\metre\squared} \\
    Label standard deviation $s$ & \qty{235921}{\ampere\per\metre\squared} \\
    Training noise levels & $\ln\sigma \sim \mathcal{N}(P_\mathrm{mean}, P_\mathrm{std}^2)$, $P_\mathrm{mean} = -0.5$, $P_\mathrm{std} = 1.5$ \\
    Training steps & \num{100000} \\
    Batch size & 64 \\
    Optimizer & AdamW~\cite{Loshchilov2019} \\
    Peak learning rate & $3 \times 10^{-4}$ \\
    Learning-rate schedule & 500 warmup steps, then cosine decay \\
    Weight decay & $10^{-4}$ \\
    Exponential moving average decay & 0.9995 \\
    Probability of zero conditioning vector & 0.1 \\
    Training time & 53~min on one RTX 5090 \\
    \bottomrule
  \end{tabular}
\end{table}

Table~\ref{tab:sampler-settings} lists the sampler settings. We eigendecompose $\Amat^\top\Amat$ once for each mask, and each correction then multiplies by the matrix of eigenvectors and by its transpose. One frame takes 1.1~s on one RTX 5090 GPU and one CPU thread, and the correction step on the CPU takes about 80\% of this time. With the same masks, one LIUQE solution takes 0.06--0.07~s on 32 CPU threads, starting from the previous frame. In the benchmark, each evaluation frame is preceded by up to 50 warm-up frames, so LIUQE takes 2.6--2.8~s per evaluation frame.

\begin{figure}[t]
  \centering
  \includegraphics[width=\textwidth]{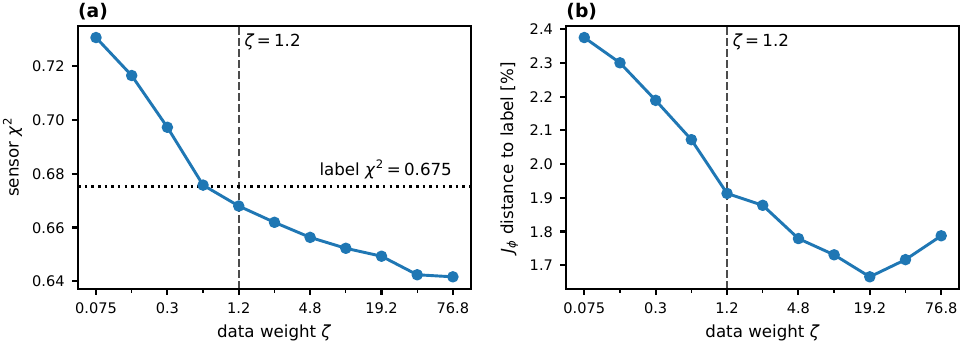}
  \caption{Selection of the data weight $\zeta$ on the 27 validation shots with all sensors: (a)~$\chi^2$ and (b)~label distance of $\jphi$.}
  \label{fig:zeta}
\end{figure}

We chose $\zeta$ on the 27 validation shots with all sensors, 270 frames in total (figure~\ref{fig:zeta}). We scanned 11 candidate values from 0.075 to 76.8. As $\zeta$ increases, we take the first value at which the $\chi^2$ of the reconstruction falls below the $\chi^2$ of the label, 0.675. The selected value is $\zeta = 1.2$. In the same scan, the label distance of $\jphi$ is 2.38\% at $\zeta = 0.075$, 1.91\% at $\zeta = 1.2$, 1.67\% at $\zeta = 19.2$ and 1.79\% at $\zeta = 76.8$. The selection rule does not use the label distance, and a larger $\zeta$ would have given a lower label distance on these shots. In this scan, the label distance at the selected value is therefore conservative.

\begin{table}[t]
  \centering
  \caption{Settings of the sampler, with $\onevec$ the vector of ones.}
  \label{tab:sampler-settings}
  \small
  \begin{tabular}{@{}l>{\raggedright\arraybackslash}p{0.46\textwidth}@{}}
    \toprule
    Setting & Value \\
    \midrule
    Sampling steps & 64 \\
    Noise levels & EDM sequence from 160 to 0.004 with $\rho = 7$, then 0~\cite{Karras2022} \\
    Samples per frame $K$ & 8 \\
    Sample exclusion & $I_\mathrm{p}$ error above 1.5 times the median over the $K$ samples \\
    Normalized coordinates & $\xvec = (\jvec - \bar{j}\,\onevec)/s$ \\
    Weight of the correction & $\lambda = 1/(\zeta(\sigma_t^2 + 10^{-4}))$ \\
    Matrix and data & $\Amat s$ and $\bvec - \bar{j}\,\Amat\onevec$ \\
    Data weight $\zeta$ & 1.2 \\
    Standard deviation of the $I_\mathrm{p}$ row & \qty{5}{\kilo\ampere} \\
    Standard deviation of rows outside the limiter & $0.02\,s$ \\
    Solver & eigendecomposition of $\Amat^\top\Amat$ once per mask, double precision \\
    Reconstruction time & 1.1~s per frame, one RTX 5090 GPU and one CPU thread \\
    \bottomrule
  \end{tabular}
\end{table}

Table~\ref{tab:label-settings} lists the LIUQE settings for the labels, where $\psi_\mathrm{A}$ and $\psi_\mathrm{B}$ are the flux at the magnetic axis and at the plasma boundary. LIUQE weights each magnetic channel by the inverse of a standard deviation from its machine description file, which differs from the $\sigma_i$ of the likelihood. In the evaluation, the $\chi^2$ of both methods is computed with the $\sigma_i$ of the likelihood.

\begin{table}[t]
  \centering
  \caption{Settings of LIUQE for the labels, with the parameter names of the MEQ toolbox. A dash marks a setting without a single parameter name.}
  \label{tab:label-settings}
  \small
  \begin{tabular}{@{}ll>{\raggedright\arraybackslash}p{0.4\textwidth}@{}}
    \toprule
    Setting & MEQ parameter & Value \\
    \midrule
    Profile basis & \texttt{bfct}, \texttt{bfp} & \texttt{bfabmex}, \texttt{[1 2]} \\
    Basis of $p'$ & -- & $\psi - \psi_\mathrm{B}$ \\
    Basis of $FF'$ & -- & $\psi - \psi_\mathrm{B}$, $(\psi - \psi_\mathrm{B})(\psi - \psi_\mathrm{A})$ \\
    Free profile coefficients & -- & 3 \\
    Vessel currents as unknowns & \texttt{ivesm}, \texttt{selu} & 1, \texttt{'v'} (104 filaments) \\
    Error of vessel currents & \texttt{Iuerr} & \qty{1}{\kilo\ampere} \\
    Error of $I_\mathrm{p}$ & \texttt{Iperr} & \qty{5}{\kilo\ampere} \\
    Error of PF coil currents & \texttt{Iaerr} & \qty{1}{\kilo\ampere}-turn \\
    Channel weight & -- & inverse of the machine-file standard deviation, 0 for 15 channels \\
    Diamagnetic loop & \texttt{idml} & 0 (not used) \\
    Initial guess & \texttt{initguess} & \texttt{'previous'} \\
    Convergence rate & -- & 95.68\% of \num{421697} frames \\
    \bottomrule
  \end{tabular}
\end{table}

The evaluation shots form three blocks, \#41733--\#41747 (LSN), \#41544--\#41550 (USN) and \#41561--\#41575 (DN). A buffer of 37 shots, which includes the eight shots on either side of each block, is left out of every set.

\section{Plasma shapes}
\label{app:shapes}

A single prior is trained on all plasma shapes of the training data, that is, lower single null, upper single null, double null and negative triangularity (sections~\ref{sec:prior} and~\ref{sec:data-eval}). For the evaluation, we divide the 370 evaluation frames into three shape classes. The LSN class has 149 frames from 15 shots, the USN class has 70 frames from 7 shots, and the DN class has 141 frames from 15 shots. Ten frames belong to no class. The LSN and USN classes are the frames of the LSN and USN blocks whose label $\psivec$ has LSN and USN topology, respectively. The DN class is the frames of the DN block, shots \#41561--\#41575, with $|dR_\mathrm{sep}| \le 3$~cm. Here $dR_\mathrm{sep}$ is the radial distance between the separatrices through the two X-points. We estimate it at the outermost point of the label boundary as $(\psi_{\mathrm{X}2} - \psi_\mathrm{B})/|\partial\psi/\partial R|$, where $\psi_{\mathrm{X}2}$ is the flux at the second X-point, and use only its magnitude. Within this DN experiment, the magnitude of the per-shot median of $dR_\mathrm{sep}$ lies above 1~cm in some shots and below 1~cm in others. The DN class therefore also includes frames whose label topology is single null. Of its 141 frames, 100 have DN topology, 39 have LSN topology and 2 have USN topology.

The frames of figure~\ref{fig:shapes} are shot \#41741 at $t = \qty{6.868}{\second}$ for LSN, shot \#41548 at $t = \qty{1.545}{\second}$ for USN and shot \#41561 at $t = \qty{1.286}{\second}$ for DN. In each class, the frame is the one with the median label distance of the diffusion reconstruction at dropout 0.9. We chose the DN frame among the frames on which both X-points of the label are active, that is, both lie on the separatrix of the label.

\begin{figure}[tbp]
  \centering
  \includegraphics{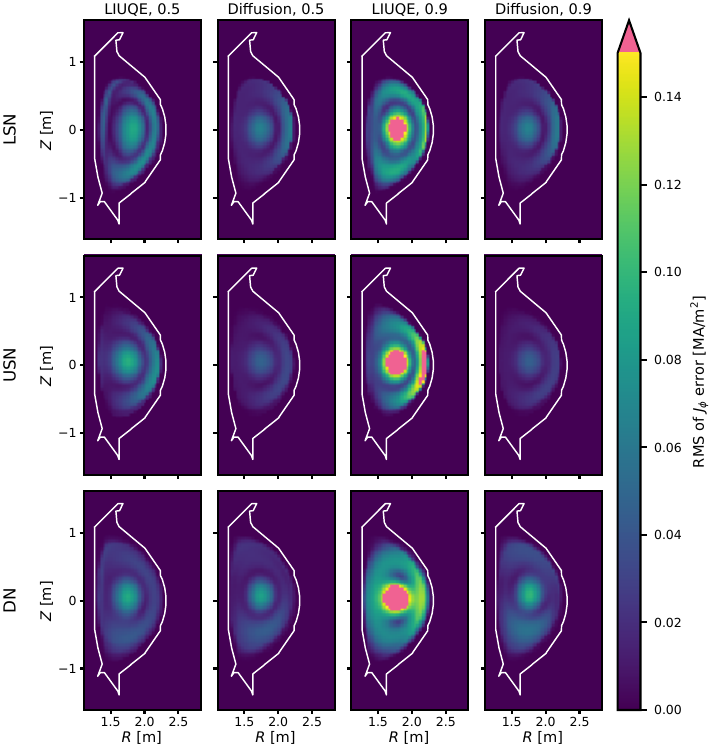}
  \caption{Error maps of $\jphi$ for LIUQE and the diffusion reconstruction at dropout 0.5 and 0.9. Rows are the LSN (149 frames), USN (70 frames) and DN (141 frames) classes, and pink marks values above the color scale.}
  \label{fig:jerr}
\end{figure}

\begin{figure}[tbp]
  \centering
  \includegraphics{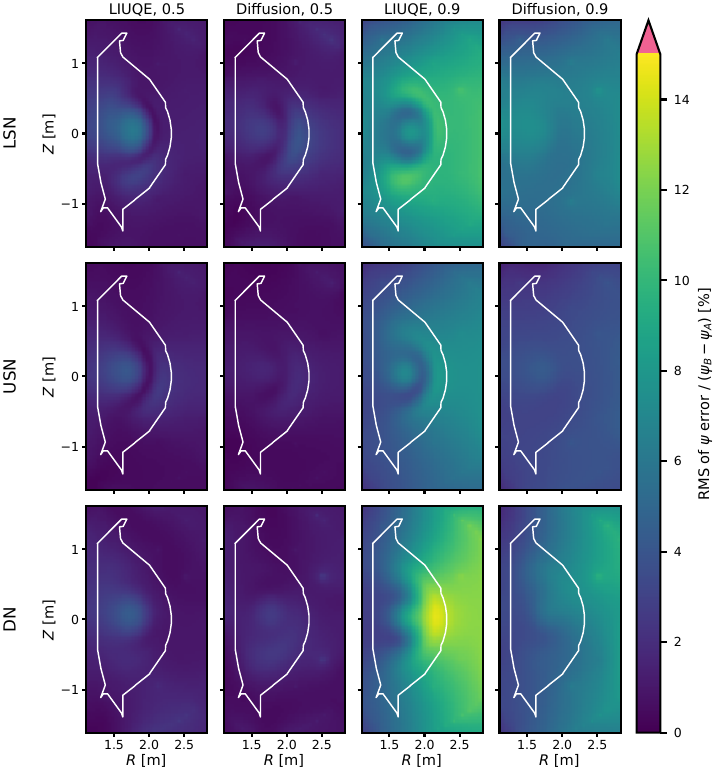}
  \caption{Normalized error maps of the poloidal flux $\psivec$, in the layout and color convention of figure~\ref{fig:jerr}.}
  \label{fig:psierr}
\end{figure}

Figure~\ref{fig:jerr} maps the error of $\jphi$ by shape class, with rows for the classes and columns for the two methods at dropout 0.5 and 0.9. At each grid point, the map shows the root mean square (RMS) of the error over the frames of the class, in \unit{\mega\ampere\per\metre\squared}. LIUQE excludes the frames on which it did not converge, and the diffusion reconstruction uses all frames of the class. The maps of $\psivec$ follow the same rule. The color scale ends at \qty{0.15}{\mega\ampere\per\metre\squared} for $\jphi$ and at 15\% for $\psivec$, and pink marks values above it. Figure~\ref{fig:psierr} maps the error of $\psivec$ in the same layout. For each frame, we compute $(\psi_\mathrm{rec} - \psi_\mathrm{label})/(\psi_\mathrm{B} - \psi_\mathrm{A})$ and take the RMS over the frames at each grid point. Here $\psi_\mathrm{rec}$ and $\psi_\mathrm{label}$ are the reconstructed and label flux, and $\psi_\mathrm{A}$ and $\psi_\mathrm{B}$ are the flux of the label at the magnetic axis and at the plasma boundary. We do not divide by the RMS of the label $\psivec$. The vacuum flux extends over the whole grid, so that denominator would change with the extent of the grid. For every shape class at dropout 0.5 and 0.9, the median of the $\psivec$ map over the grid points inside the limiter is lower for the diffusion reconstruction than for LIUQE. At dropout 0.9, this median is 6.11\% against 7.77\% for LSN, 3.27\% against 5.01\% for USN and 4.48\% against 6.38\% for DN (diffusion reconstruction against LIUQE). At dropout 0.9, the error spreads smoothly over the whole grid. This pattern is consistent with section~\ref{sec:vessel}, where most of the dropout dependence of the error of $\psivec$ comes from the estimate $\hat{\Ivec}_\mathrm{v}$ of the vessel currents.

\section{Calibration of the sample spread}
\label{app:spread}

This appendix defines the statistics of section~\ref{sec:uncertainty} and reports the rest of them. The reported reconstruction is the mean of the $K$ samples. At each pixel of a frame, let $j^{(k)}$ be the values of the samples of $\jphi$ that remain after the samples with a large $I_\mathrm{p}$ error are excluded, and let $j_\mathrm{label}$ be the value of the label. We define
\begin{equation}
  z = \frac{j_\mathrm{label} - \mu}{s_K}, \qquad \mu = \frac{1}{K}\sum_{k=1}^{K} j^{(k)}, \qquad s_K^2 = \frac{1}{K-1}\sum_{k=1}^{K} \big(j^{(k)} - \mu\big)^2,
  \label{eq:zscore}
\end{equation}
at each pixel where $j_\mathrm{label} \neq 0$ and $s_K > 0$. In frames where samples were excluded, $K$ in equation~\eqref{eq:zscore} is the number of remaining samples. If the samples and the label are drawn independently from the same Gaussian predictive distribution, the spread is calibrated and $z$ follows $\sqrt{1+1/K}\,t_{K-1}$, where $t_{K-1}$ is the Student $t$ distribution with $K-1$ degrees of freedom. For $K = 8$, the median of $|z|$ is then 0.754, which we call the calibrated value. It is an approximation in frames where samples were excluded. The coverage of a nominal interval of level $\alpha$ is the fraction of these pixels with $|z| < \sqrt{1+1/K}\,q_{K-1}\big((1+\alpha)/2\big)$, where $q_{K-1}(p)$ is the $p$ quantile of $t_{K-1}$. Interval coverage is a necessary but not sufficient check of a posterior estimator~\cite{Lemos2023}. We compute each statistic per frame and report the median over the frames of each shot and then over the 37 evaluation shots.

The sample spread of $K = 8$ samples is not a calibrated uncertainty. In the ten settings, the nominal 95\% interval contains the label at 43--48\% of the pixels. The nominal 68\% interval contains it at 18--21\%. The median of $|z|$ is 3.5--4.2 times the calibrated value 0.754.

Where the spread is larger, the error is also larger, but the spread does not predict the size of the error. Here the error is $|j_\mathrm{label} - \mu|$ at a pixel, and the ratio of error to spread is the median over pixels of $|j_\mathrm{label} - \mu|/s_K$. We pool the pixels of all frames of a setting and group them into deciles of $s_K$. From the first decile to the last, the median error rises, but the ratio of error to spread falls. At dropout 0 it is 3.62 in the first decile and 1.92 in the last. In all ten settings, the ratio in the first decile is larger than in the last.

The underestimate of the error by the spread does not come from the choice of $\zeta$. We computed $z$ from the samples of the $\zeta$ scan of appendix~\ref{app:settings}, which uses the validation shots with all sensors. In this scan, the median of $|z|$ does not fall below 1.9 times 0.754. The median of $|z|$ is smallest at the selected value, $\zeta = 1.2$.

\clearpage
\bibliographystyle{unsrtnat}
\bibliography{ref}

\end{document}